\documentclass[prb,showpacs,a4paper,twocolumn]{revtex4-1}%

\usepackage{amsmath}
\usepackage{graphicx}

\DeclareMathOperator{\Tr}{Tr}

\newcommand{\kB}{k_{\mathrm{B}}}
\newcommand{\SB}{S_{\mathrm{B}}}
\newcommand{\TB}{T_{\mathrm{B}}}
\newcommand{\Amu}{A_\mu}
\newcommand{\fix}[2]{{#2}\Big|_{#1}}
\newcommand{\SG}{S_{\mathrm{G}}}
\newcommand{\TG}{T_{\mathrm{G}}}
\newcommand{\barSG}{{\bar S}_{\mathrm{G}}}

\newcommand{\Estar}{{E^\star}}
\newcommand{\Emax}{E_M}

\newcommand{\eps}{\epsilon}

\newcommand{\latin}[1]{{\itshape #1}}
\newcommand{\french}[1]{{\itshape #1}}

\newcommand{\cf}{\latin{cf.}}
\newcommand{\eg}{\latin{e.\,g.}}
\newcommand{\ie}{\latin{i.\,e.}}
\newcommand{\etal}{\latin{et al.}}

\newcommand{\via}{\latin{via}}

\newcommand{\mutmut}{\latin{mutatis mutandis}}

\newcommand{\ala}{\french{\`a la}}

\newcommand{\Eqref}[1]{Eq.~\eqref{#1}}
\newcommand{\Eqsref}[1]{Eqs.~\eqref{#1}}

\newcommand{\Refcite}[1]{Ref.~\onlinecite{#1}}

\begin{document}

\title{Gibbs, Boltzmann, and negative temperatures} 

\author{Daan Frenkel} 

\affiliation{Department of Chemistry, University of Cambridge,
  Lensfield Road, Cambridge, CB2 1EW, U.K.}
  
\author{Patrick B. Warren} 
  
\affiliation{Unilever R\&D Port Sunlight, Quarry Road East, Bebington,
  Wirral, CH63 3JW, U.K.}

\pacs{05.20.Gg, 05.70.-a}


\date{May 16, 2014 -- conditionally accepted version, Am.~J.~Phys.}

\begin{abstract}
In a recent paper, Dunkel and Hilbert [Nature Physics {\bf 10}, 67--72
  (2014)] use an entropy definition due to Gibbs to provide a
``consistent thermostatistics'' which forbids negative absolute
temperatures.  Here we argue that the Gibbs entropy fails to satisfy a
basic requirement of thermodynamics, namely that when two bodies are
in thermal equilibrium, they should be at the same temperature.
The entropy definition due to Boltzmann does meet this test, and
moreover in the thermodynamic limit can be shown to satisfy Dunkel and
Hilbert's consistency criterion.  Thus, far from being forbidden, negative
temperatures are inevitable, in systems with bounded energy
spectra.
\end{abstract}

\maketitle 

The concept of negative temperature is one which has often been
discussed in the pages of this Journal.\cite{Tyk75,
  Dan76, Tyk76, Tre76, Tyk78} There has recently been renewed interest
in the area.\cite{Mos05, ACP08, RMR10, BRS13, Rom13, DH14} Formally, a
negative temperature corresponds to a system under conditions in which
the number of states $\omega(E)$ in the vicinity of energy $E$ is a
decreasing function, to wit $1/T\equiv \partial S/\partial E<0$ where
$S=\kB\ln\omega$ is the Boltzmann entropy. This situation usually
arises in a system with a bounded energy spectrum.  It has been
realized experimentally in famous early work on nuclear spin systems
by Purcell and Pound,\cite{PP51} and in very recent experiments on
ultra-cold quantum gases by Braun \etal\cite{BRS13} The existence of
negative temperatures leads naturally to many questions.  Prominent
amongst these is whether a Carnot cycle can be constructed with a
negative temperature reservoir, and whether the consequent Carnot
efficiency can be greater than one.\cite{Tyk78, Zem57, SS74, Lan77}

In a recent article\cite{DH14} Dunkel and Hilbert argued that the use
of Boltzmann's definition of entropy in the above expression for the
temperature leads to unphysical predictions.\cite{Planck} They propose
instead to define entropy using Gibbs' notion of ``extension in phase
space'' (see below). \cite{Gib02} They argue that this approach, which
they attribute to Gibbs,\cite{Note1} cannot give rise to negative
temperatures nor can it predict the existence of Carnot cycles with an
efficiency larger than one.  Moreover, they show that the Boltzmann
approach gives rise to unphysical predictions for systems with very
few degrees of freedom.  In what follows, we will ignore the latter
point. As Gibbs himself stated on several occasions, it is
unreasonable to expect a meaningful correspondence between statistical
mechanics and thermodynamics for systems with only a few degrees of
freedom.  Our discussion will focus only on systems with very many
degrees of freedom, \ie\ those which are in the proverbial
``thermodynamic limit''.\cite{Kuz14}

The point about negative temperatures and Carnot cycles is more
interesting.  We shall argue below that Dunkel and Hilbert are
mistaken in their attempt to deprecate the Boltzmann entropy, and that
instead it is the Gibbs entropy which fails to meet a basic criterion
of thermodynamics.  In contrast, the Boltzmann entropy does meet this
basic test, and we also prove that it satisfies Dunkel and Hilbert's
consistency criterion in the thermodynamic limit.  With the Boltzmann
entropy, negative temperatures are inevitable in systems with bounded
energy spectra, but this is not a problem if one pays attention to the
details as explained by Ramsey over half a century ago.\cite{Ram56} To
round off we present a pedagogical example in which we construct a
concrete example of a Carnot cycle connecting reservoirs of opposite
temperatures, thereby exhibiting a Carnot efficiency bigger than one.

\section{Critique of the Gibbs entropy}
Let us briefly define the key quantities in modern notation. In
particular, we will replace Gibb's ``extension in phase space'' by
$\Omega(E)$, the total number of quantum states of a system with an
energy less than or equal to $E$.  For classical systems with many
degrees of freedom, $\Omega(E)$ is dominated by the number of states
very close to $E$. To compute the number of states in a narrow
interval $\Delta\epsilon$ around $E$, we simply differentiate
$\Omega(E)$ with respect to $E$, to obtain
\begin{equation}
\frac{\partial\Omega(E)}{\partial E}\,\Delta\epsilon 
\equiv \omega(E)\,\Delta\epsilon\,.\label{eq:Oo}
\end{equation}
Boltzmann's definition of entropy is 
\begin{equation}
\SB(E)=\kB\ln\omega(E) +\text{constant}\,.
\end{equation}
In contrast, Gibbs also considered other definitions of entropy, including
\begin{equation}
\SG(E)=\kB\ln\Omega(E)+\text{constant}\,.
\end{equation}
Of course, Gibbs did not include $\kB$ in his definition but, again,
that is immaterial for the remainder of our argument.  In what
follows, we leave out the additive constant, as it is irrelevant for a
discussion of thermal equilibrium and heat engines.  

The definition of temperature in thermodynamics is
\begin{equation}
T= \Bigl(\frac{\partial S(E)}{\partial E}\Bigr)^{-1}\,.
\end{equation}
For classical systems with many degrees of freedom, the difference in
the value of the temperature based on $\SB$ and $\SG$ is negligible; the
reason being that for such systems $\omega(E)$ increases very steeply
with $E$, hence a constraint $\epsilon\le E$ is almost equivalent with
$\epsilon=E$. However, for systems with an energy that is bounded from
above, $\omega(E)$ may decrease for large energies, whereas
$\Omega(E)$ is monotonically increasing. An example is a system of $N$
non-interacting spins, which we shall discuss in more detail below. In
the regime where $\omega(E)$ is not a monotonically increasing
function of $E$, the two definitions of entropy lead to very different
results for the temperature of a macroscopic system (one negative, the
other positive). Only one can be right. It turns out that, in contrast
to what is argued in \Refcite{DH14}, to meet a basic
requirement of thermodynamics we must use the definition based on
Boltzmann's entropy $\SB(E)$.

The nub of our argument turns on the behavior of systems, considered
jointly, which are able to exchange energy.  We shall argue that the
condition for these to be in thermal equilibrium is that they all have
the same temperature \ala\ Boltzmann.  This is essentially what is
often termed ``the zeroth law of thermodynamics''.  It implies that the
Boltzmann entropy, \emph{and only that}, can be used to construct a
universal thermodynamic temperature scale.  Without this basis, the
whole edifice of classical thermodynamics is built on sand.

To see this, let us briefly restate the key properties that the
statistical mechanical entropy and temperature should reproduce in
order to correspond with the thermodynamic quantities defined by
Clausius:
\begin{enumerate}
\item if a closed system is in equilibrium, its entropy must be at a
  maximum;
\item heat never spontaneously flows from cold to hot
\cite{Note2}
\item at the end of one cycle of an reversible heat engine the entropy of
  the system (\ie\ engine plus temperature reservoirs) has not
  changed.
\end{enumerate}
It should be stressed that Clausius' thermodynamics is based on
experimental observations, not on axioms.  The simplest way to
construct statistical mechanics is to invoke the so-called {\em
  ergodic hypothesis}: ``A system at a given energy $E$ is equally
likely to be found in any of its $\omega(E)$ quantum
states''.~\cite{Note3}  The ergodic hypothesis does {\em not} apply to
$\Omega(E)$. If a system is at a fixed energy $E$, it does not visit
states with a lower energy.  These lower-energy states are therefore
irrelevant for counting the degeneracy of the system.

Let us consider a closed system with total energy $E$, comprised of
two subsystems with energies $E_1$ and $E_2$. We assume that the
absolute value of the interaction energy between the two subsystems is
much smaller than $E$. In that case, $E=E_1+E_2$. The total number of
states of this system is given by
\begin{equation}
\omega_T(E_1,E_2)=\omega_1(E_1)\times \omega_2(E_2)\,.
\end{equation}
Of course, we can write the same expression for $\Omega(E_1,E_2)$ but,
importantly, if we consider all the states for which $E_1+E_2\le E$,
we are {\em not} describing a system at fixed energy $E$, but rather a
system with a fixed maximum energy $E$.\cite{Note3.5} For classical
systems, this distinction is irrelevant in the thermodynamic limit,
but for a system with an energy that is bounded from above (say
$E\le\Emax$), the difference is crucial.  To illustrate this, let us
consider an extreme (but perfectly legitimate) example: suppose that
we regard $E_2<\Emax$ as merely a cap on the energy of system $2$.
Then, increasing $E_2$ by an amount $\Delta E$ does not necessarily
correspond to an energy transfer of $\Delta E$ from system $1$ to
system $2$: there is no relation between $\Delta E$ and heat transfer
and all relation with thermodynamics is lost.  To establish a link
with thermodynamics it is essential that the total energy $E$ is equal
to the sum $E_1+E_2$. This condition can only be enforced if we work
with subsystems that have well-defined energies, not just systems with
an energy below a particular value.

Let us next consider what happens if we allow energy exchange between
systems 1 and 2. In equilibrium, the entropy should be at a maximum
with respect to a small energy transfer from system 1 to system
2. Hence
\begin{equation}
\frac{\partial \ln\omega_T}{\partial E_1}=0\,.
\end{equation}
Using the fact that $dE_1=-dE_2$ we obtain
\begin{equation}
\frac{\partial \ln\omega_1(E_1)}{\partial E_1}=
\frac{\partial \ln\omega_2(E_2)}{\partial E_2}
\end{equation}
and hence
\begin{equation}
\frac{1}{T_1}=\frac{1}{T_2}\,.\label{eq:zeroth}
\end{equation}
Note that we obtain this result only if we use the Boltzmann entropy
$S=\kB\ln\omega$.  

But, should the Gibbs entropy be so lightly dismissed?  After all, as
Dunkel and Hilbert point out, there is a mathematically rigorous
equipartition theorem (Eq.~(8) of \Refcite{DH14}) which
features $\SG$ and would appear to single out the associated
temperature $\TG=(\partial\SG/\partial E)^{-1}$ as a privileged
quantity.  However, it is dangerous to extrapolate from this that
equality of $\TG$ can be used to gauge whether systems are in thermal
equilibrium; this can easily lead to absurd results.  For example,
consider the finite-size spin system in Fig.~1 of
\Refcite{DH14}.  In the population-inverted state (see below
for the definition) the Boltzmann temperature is negative but $\TG$ is
positive and finite.  One can easily construct a classical system (for
example a perfect gas) with the same value of $\TG$.  Should one
therefore conclude that the spin system with an inverted population
can be in thermal equilibrium with a perfect gas?  Of course not!  The
spin system would lose energy to the gas raising the (Boltzmann) entropy
of both.  In the conventional picture of course, a population
inverted state has a negative Boltzmann temperature and is always
``hotter'' than a normal system with a positive Boltzmann temperature,
so one would always expect heat transfer to take place.

The only remaining point to discuss is Dunkel and Hilbert's
consistency criterion.  In \Refcite{DH14} it is proved that
$\SG$ satisfies this requirement, independent of system size.  Dunkel
and Hilbert further argue that $\SB$ must therefore fail this
requirement because typically $\SB\ne\SG$ for small systems.  However
it can be proved that in the thermodynamic limit $\SB$ \emph{does}
satisfy the consistency criterion.  In Appendix \ref{sec:appsb} we
present such a proof.  Our conclusion therefore is that $\SB$ meets
all the requirements one would expect of a thermodynamic entropy,
whereas $\SG$ does not.

\begin{figure}
\begin{center}
\includegraphics[clip=true,width=2.2in]{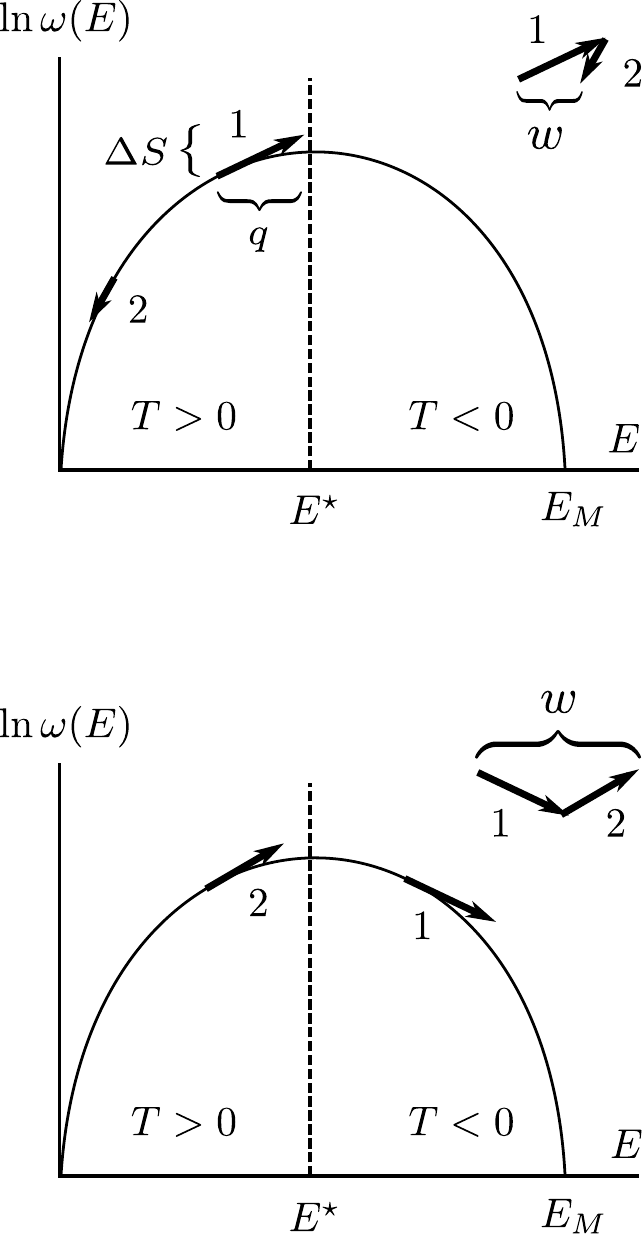}
\end{center}
\caption{If a reversible heat engine operates between reservoirs $1$
  and $2$ that both have positive temperatures (top panel; $T_1>T_2$),
  then the condition that $S$ is constant implies that as heat flows
  out of reservoir $1$, heat must flow into reservoir $2$.  However,
  if a reversible heat engine operates between reservoir ($1$) at a
  negative temperature and another ($2$) at positive temperature
  (bottom panel), then the condition that $S$ is constant implies that
  as heat flows out of reservoir $1$, heat must also flow out of
  reservoir $2$.  The curves in these figures represent the entropy of
  the engine.  Since $\Delta S=q/T$ and $1/T=\partial S/\partial E$,
  it follows that $\partial S/\partial E=\Delta S/q$ and the arrows
  have a specific geometric interpretation: the horizontal
  displacement is the heat gain $q$ and the vertical displacement is
  the entropy change $\Delta S$.  The Carnot cycle itself can then be
  represented by combining the arrows vectorially since in the
  adiabatic steps $q=\Delta S=0$.  The resultant vector should have no
  net vertical displacement since the net entropy change should
  vanish; the net horizontal displacement represents $w$, the amount
  of energy available to do useful work.  This graphical construction
  is shown in the insets in the main panels.  It demonstrates why the
  Carnot efficiency $\eta\equiv w/q_1<1$ in the top panel, and
  $\eta>1$ in the bottom panel.\label{fig:01}}
\end{figure}

\section{Negative temperatures and Carnot cycles}
Now that we have convinced ourselves we should use the entropy
\ala\ Boltzmann, the implication is that, if $\omega(E)$ is not a
monotonically increasing function, negative temperatures are
inevitable.  Are negative temperatures a problem? Not really. First of
all, heat still only flows from ``hot'' (low $1/T$) to ``cold'' (high
$1/T$). But indeed, Carnot efficiencies can be larger than
one.\cite{Note4} This may seem strange, but it violates no known law
of nature.\cite{Note4.1}

Let us consider a Carnot cycle operating between two heat reservoirs
(see Fig.~\ref{fig:01}).  The ``hot'' reservoir 1 operates at a
negative temperature $T_1<0$. The ``cold'' reservoir 2 operates at a
positive temperature $T_2>0$.  We now operate an engine between the
two reservoirs. The first law of thermodynamics states that the work
$w$ done by the engine must be equal to the heat $q_1-q_2$ absorbed by
the engine (energy conservation). It is important to look at the sign
definitions: a positive sign of $q_1$ means that heat flows {\em out
  of} reservoir 1. A positive sign of $q_2$ means that heat flows {\em
  into} reservoir 2.  Each stage in the cycle is reversible so that
the entropy changes of the engine are respectively $\Delta S_1=
q_1/T_1$ and $\Delta S_2=-q_2/T_2$.  Perhaps confusingly, $\Delta
S_1<0$, but this must be so because when heat flows out of a negative
temperature reservoir, its entropy increases.  Now, since $S$ is a state
function, the total entropy of the engine does not change, and $\Delta
S_1+\Delta S_2=0$.  This implies
\begin{equation}
\frac{q_1}{T_1}=\frac{q_2}{T_2}\,.
\end{equation}
If $T_1$ is negative and $T_2$ is positive, then the signs of
$q_1$ and $q_2$ must be opposite. Hence, if heat flows {\em out of}
reservoir 1 ($q_1>0$), then reversibility of the engine requires that
heat {\em also flows out of reservoir 2} ($q_2<0$).  In such a
situation, the work done is
\begin{equation}
w = q_1-q_2 = q_1+|q_2|\,.
\end{equation}
The efficiency of the Carnot cycle is defined as the ratio 
\begin{equation}
\eta=\frac{w}{q_1}=1+\frac{|q_2|}{q_1}>1\,.
\end{equation}
Hence, indeed, the Carnot efficiency is larger than one. However, no
physical law forbids that. There is one potential point of concern: it
would seem that it is possible to run an engine in contact with a
single, negative-temperature heat bath.  Whilst this is true, negative
temperature heat baths do not occur naturally: they have to be
prepared (in the case of lasers, we call this ``pumping'').
If the total heat and work budget includes the preparation of the
negative-temperature heat bath from a system at positive temperatures,
then it turns out that it is still not possible to run an engine
sustainably by extracting heat from a single reservoir.  Negative
temperatures do not imply perpetual motion.

\section{Spin systems}
Spin systems form an excellent arena to explore the issues raised
here.  Apart from the three-level system, much of this material can be
found in the original literature,\cite{Ram56} and
textbooks.\cite{Zem57, Wal85} For an interesting mechanical analog of
these systems, see \Refcite{ACP08}.

\subsection{Two level system}
The simplest case to consider is a system of $N_0$ spins in a ground
state with zero energy and $N_1$ spins are in an excited state at
energy $\eps$, such that the total number of spins $N=N_0+N_1$ is
fixed.  Often one can visualize the $N_0$ spins as ``down'' and the
$N_1$ spins as ``up'', but care is needed with this mental picture as we
shall consider cases where $\eps<0$ and $N_1>N_0$ (population
inversion).  We shall suppose that the spins are distinguishable so
that the number of states which can accommodate this arrangement of
spins is $\omega = N!/(N_0!\,N_1!)$.  For this system, the energy and
(Boltzmann) entropy are therefore, respectively,
\begin{equation}
\begin{split}
E&=\eps N_1\,,\\[3pt]
S&=\ln\omega=N\ln N-N_0\ln N_0-N_1\ln N_1\,.
\end{split}
\label{eq:es}
\end{equation}
We have supposed $N\gg1$ in the entropy expression.  For notational
simplicity we set $\kB=1$ and drop the subscript ``B'' from $\SB$ since
we shall be exclusively considering the Boltzmann entropy.  

Since $N_0=N-N_1$, and $N$ is fixed, the (Boltzmann) temperature of
the spin system is
\begin{equation}
\frac{1}{T}=\frac{\partial S}{\partial E}
=\frac{(\ln N_0+1)-(\ln N_1+1)}{\eps}
=\frac{1}{\eps}\ln\frac{N_0}{N_1}\,.
\end{equation}
Therefore one obtains the familiar Boltzmann result
\begin{equation}
{N_1}={N_0}\,e^{-\eps/T}\,.\label{eq:boltz}
\end{equation}
It follows that
\begin{equation}
\frac{N_1}{N}=\frac{e^{-\eps/T}}{1+e^{-\eps/T}}\,,\quad
\frac{N_0}{N}=\frac{1}{1+e^{-\eps/T}}\,.
\label{eq:eos}
\end{equation}
The free energy $F=E-TS$ has to be expressed in terms of its ``natural''
variables, $T$ and $N$.  We first have
\begin{equation}
\begin{split}
S&=N\ln N-N_0\ln N_0-N_1\ln N_1\\[3pt]
&=(N_0+N_1)\ln N-N_0\ln N_0-N_1\ln N_1\\[3pt]
&=-N_0\ln(N_0/N)-N_1\ln(N_1/N)\,.
\end{split}
\end{equation}
Consequently,
\begin{equation}
F=\eps N_1+TN_0\ln\frac{N_0}{N}+TN_1\ln\frac{N_1}{N}\,.
\end{equation}
We substitute \Eqsref{eq:eos} into this to find
\begin{equation}
F=-TN\ln(1+e^{-\eps/T})\,.\label{eq:f}
\end{equation}
This rather neat result follows after a few lines of algebra which is
left as an exercise for the reader.  It can also be derived from the
partition function sum, which is left as a further exercise.

From \Eqref{eq:f}, the quantity conjugate to $\eps$ is
\begin{equation}
-\frac{\partial F}{\partial\eps} = 
-\frac{N\,e^{-\eps/T}}{1+e^{-\eps/T}}=-N_1\,.
\end{equation}
This makes $N_1$ the natural variable to describe the arrangement of
the spins.  For this system, the $N_1$-$\eps$ plane is the analog of
the $p$-$V$ diagram encountered in textbooks.\cite{Note4.2} The first
expression in \Eqsref{eq:eos} serves as an ``equation of state'',
providing isotherms in the $N_1$-$\eps$ plane.

\begin{figure}
\begin{center}
\includegraphics[clip=true,width=2.6in]{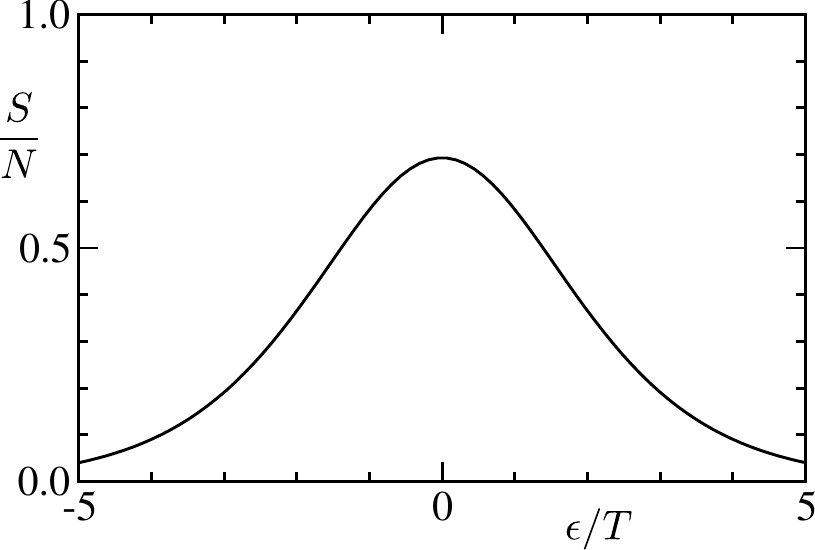}
\end{center}
\caption{The entropy function of \Eqref{eq:s2}.\label{fig:02}}
\end{figure}

For completeness, the entropy as a function of $T$ and $N$ follows
most simply by solving $F=E-TS$.  Substituting the relevant
expressions, one finds (another exercise!)
\begin{equation}
\frac{S}{N}=\frac{\eps}{T}\,\frac{e^{-\eps/T}}{1+e^{-\eps/T}}
+\ln(1+e^{-\eps/T})\,.\label{eq:s2}
\end{equation}
This function (Fig.~\ref{fig:02}; \cf\ Fig.~14.23 in
\Refcite{Zem57}) only depends on the ratio $\eps/T$ and is in
fact symmetric about $\eps/T=0$, although this is not immediately
obvious.

If we examine the derivations, we see that all of the above holds for
positive and negative $\eps$, and more crucially for positive and
negative $T$ too.  If $\eps>0$ and $T>0$ the spin population is
``normal'' in the sense that the majority of spins are in the ground
state, whereas if $\eps>0$ and $T<0$ the spin population is ``inverted''
in the sense that the majority of spins are in the excited state.
An analogous situation holds, \mutmut, if $\eps<0$.

These inverted states are well defined and correspond to a total
energy which is closer to the absolute maximum energy, $\Emax=\eps N$,
than to $E=0$, as we now demonstrate.  Since $E=\eps N_1$, $\Emax=\eps
N$, and $N_0=N-N_1$, the entropy can be written as
\begin{equation}
\frac{S}{N}=-\Bigl(1-\frac{E}{\Emax}\Bigr)\ln\Bigl(1-\frac{E}{\Emax}\Bigr)
-\frac{E}{\Emax}\ln\frac{E}{\Emax}\,.
\end{equation}
This function has the non-monotonic shape shown in
Fig.~\ref{fig:01} (\cf\ Fig.~14.22 in \Refcite{Zem57}),
with a maximum at $\Estar=\Emax/2$.  Therefore the spin system has a
negative temperature for $E>\Emax/2$.  The maximum in $S(E)$
corresponds to $1/T=0$, which suggests that it should be impossible to
pass adiabatically (\ie\ at constant $S$) from positive to negative
temperature through $1/T=0$.  This is in fact a general
result.\cite{SS74, Tre76}

\begin{figure}
\begin{center}
\includegraphics[clip=true,width=2.4in]{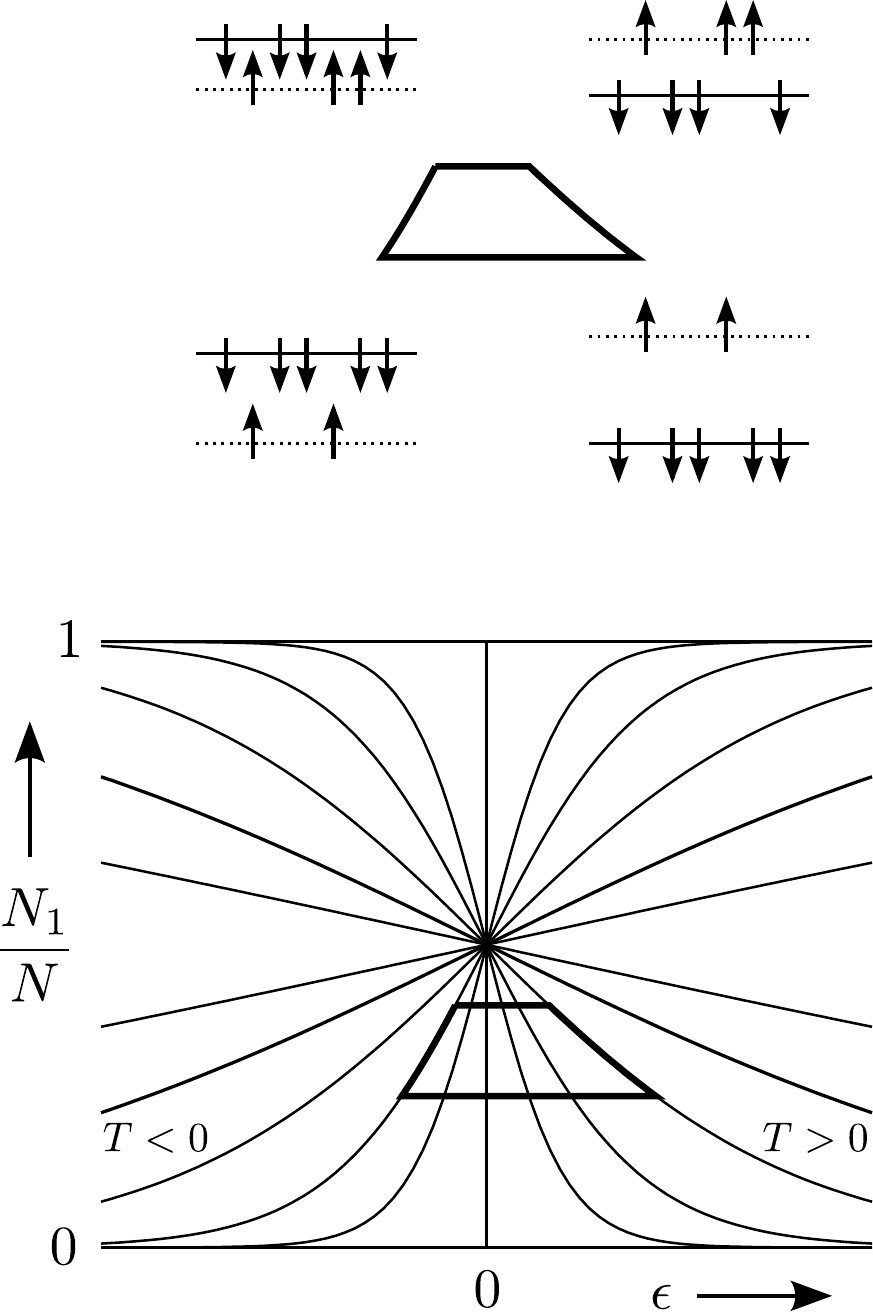}
\end{center}
\caption{A Carnot cycle connecting isotherms of opposite temperatures.
  The thin lines in the $N_1$-$\eps$ plane (lower plot) are isotherms,
  from the first expression in \Eqsref{eq:eos}.\label{fig:03}}
\end{figure}

Isotherms for the spin system in the $N_1$-$\eps$ plane are
illustrated in Fig.~\eqref{fig:03}, for both positive and negative
temperatures.  The spin system can be moved along an isotherm by
connecting it to a thermal reservoir and changing $\eps$.  The thermal
reservoir could be, for example, a much larger spin system, which can
be at a positive or negative temperature.

Adiabatic changes in the spin system correspond to changing $\eps$
without changing the distribution of spins.  One can see from
\Eqref{eq:es} that keeping $N_0$ and $N_1$ fixed leaves the entropy
$S$ unchanged.  Adiabats in the $N_1$-$\eps$ plane in
Fig.~\eqref{fig:03} are therefore horizontal lines.  One further
point can be made.  Since $N_0$ and $N_1$ depend on the ratio
$\eps/T$, it follows that $T$ remains strictly proportional to $\eps$
in an adiabatic change (\cf\ \Eqref{eq:s2}).  This observation implies
that we can invert the temperature of the spin system adiabatically,
by changing the sign of $\eps$.

\subsection{Three level system}
The three level system is worth considering because it has some novel
features not encountered in the two level system, and it sheds further
light on the problems associated with the Gibbs entropy.  The main new
feature is that a three level spin system has an internal
(macroscopic) degree of freedom, since specifying the energy and total
number of spins is insufficient to fix all the population levels.  We
start by noting the analog of \Eqref{eq:es}:
\begin{equation}
\begin{split}
N&=N_0 + N_1+ N_2\,,\\[3pt]
E&=\eps_1 N_1+\eps_2 N_2\,,\\[3pt]
S&=N\ln N-N_0\ln N_0-N_1\ln N_1-N_2\ln N_2\,.
\end{split}
\label{eq:es3}
\end{equation}
We have now explicitly included the constraint on the total number of
spins.  The entropy in this is Boltzmann's $S=\ln\omega$.  There is
absolutely no disagreement that the most likely macroscopic
arrangement of spins is the one which maximizes $\omega$, and hence
$S$.  To solve this, the usual (textbook) method is to introduce
Lagrange multipliers for the constraints (\ie\ $\mu$ for $N$, and
$\beta$ for $E$).  However the problem is sufficiently simple that one
can also proceed, for example, by eliminating $N_1$ and $N_2$ from the
above, to find $S$ as a function of $N_0$ alone, which is then
maximized.  This is certainly something which everyone should try, at
least once!  A slightly more elegant method is to differentiate all of
the above to get
\begin{equation}
\begin{split}
dN&=dN_0 + dN_1+ dN_2=0\,,\\[3pt]
dE&=\eps_1\,dN_1+\eps_2\,dN_2=0\,,\\[3pt]
dS&=-(\ln N_0+1)\,dN_0-(\ln N_1+1)\,dN_1\\
&\hspace{10em}{}-(\ln N_2+1)\,dN_2\,.
\end{split}
\label{eq:es3a}
\end{equation}
We eliminate $dN_1$ and $dN_2$ between these as though they are
algebraic quantities to find
\begin{equation}
dS=-\Bigl(
\ln N_0
+\frac{\eps_2\ln N_1}{\eps_1-\eps_2}
+\frac{\eps_1\ln N_2}{\eps_2-\eps_1}\Bigr)\,dN_0\,.
\label{eq:es3b}
\end{equation}
The coefficient in here is precisely $\partial S/\partial N_0$ which
would be obtained from the more cumbersome approach.  The condition
that $S$ is a maximum corresponds to requiring $\partial S/\partial
N_0 = 0$.  After a small rearrangement, this leads to
\begin{equation}
\frac{1}{\eps_1}\ln\frac{N_1}{N_0}=
\frac{1}{\eps_2}\ln\frac{N_2}{N_0}\,.
\label{eq:es3c}
\end{equation}
Let us pause for a moment to see what this is telling us.  An
experimentalist, for instance, would note that measurement of the
relative population of one level (\eg\ $N_1/N_0$) can be used to make
a prediction of the relative population of the other level.  Also, if
there is a population inversion for one level (\eg\ $N_1/N_0>1$ when
$\eps_1>0$), there must be population inversion for the other level
too.  Most likely though, the experimentalist will complain that we
have dressed a standard result up in rather unfamiliar language.  In
\Eqref{eq:es3c} let us call the quantity on either side of the
equality ``$-1/T$'' (or ``$-1/\kB T$'' if we wish to retain the Boltzmann
constant).  Then, \Eqref{eq:es3c} can be written
\begin{equation}
N_1=N_0\,e^{-\eps_1/T}\,,\quad N_2=N_0\,e^{-\eps_2/T}\,.
\label{eq:es3d}
\end{equation}
In other words, the populations satisfy the Boltzmann distribution.
The actual value of $T$ is determined by the constraints on $N$ and
$E$.  The case where $T<0$ corresponds to population inversion and
will inevitably arise if $E$ is made big enough at fixed $N$.
Measuring the relative population of one level is the same as
measuring $T$.  When $T$ is known, the relative population of the
other level can be predicted.  It is easy to see that this generalizes
to multilevel systems (this is where the Lagrange multiplier method
comes into its own), and it matters not one jot or tittle whether $T$
is positive or negative.

But, what happens to the Gibbs entropy and $\TG$ in this problem?
Recall that we are concerned with large systems where in the normally
populated state there is no difference between $\TG$ and $T$.  In the
population inverted state however, $\TG$ diverges exponentially with
system size,\cite{VR14} and has a complicated dependence on
the Boltzmann temperature as a function of energy (see Appendix
\ref{sec:apptg}).  From an experimental point of view this makes $\TG$
pretty much useless.

Note that in the discussion thus far, the three level spin system has
been kept isolated (\ie\ in the micro-canonical ensemble) and no
reservoirs have been involved.  We therefore cannot be accused of
introducing an assumption of ensemble equivalence by sleight-of-hand.
But we feel obliged to point to one more piece of evidence to support
the ascendancy of the Boltzmann entropy.  Specifically, once $T$ is
measured, the spin system is assured of being in thermal equilibrium
with a heat reservoir at that same temperature, irrespective of the
sign of $T$.

\subsection{Carnot cycle with efficiency greater than one}
We now return to the two level system and use our knowledge of the
thermodynamics outlined above to construct a Carnot cycle which
operates between positive and negative temperature reservoirs.  The
cycle is shown the thick solid line in the $N_1$-$\eps$ plane in
Fig.~\ref{fig:03}. The corresponding changes in the spin
population are illustrated above the main plot.  Our construction can
be viewed as a generalization of the adiabatic demagnetization
procedure widely used as a refrigeration method in experimental low
temperature physics.\cite{Zem57, Wal85}

Beginning with the upper left corner, and moving counterclockwise, the
cycle starts with a spin system in an inverted state with $\eps<0$ in
contact with a ``hot'' reservoir at some temperature $T_1<0$.  The first
step consists in isothermally increasing $|\eps|$.  The entropy of the
spin system \emph{decreases} in this step by some finite amount
$\Delta S$ (\cf\ Fig.~\ref{fig:02}, noting that $\eps/T>0$
increases).  Since we are proceeding reversibly, the entropy decrease
of the spin system is balanced by an entropy increase in the
reservoir.  Because of the peculiar properties of the negative
temperature ``hot'' reservoir, this means that the spin system withdraws
an amount of heat $q_1=|\Delta S/T_1|$ from this reservoir.  Next,
contact with the reservoir is removed and the spin population is
inverted adiabatically by reversing the sign of $\eps$ to arrive at
the lower right corner.  In the third step, the now normally populated
spin system is brought into contact with the ``cold'' reservoir at
$T_2>0$, and $\eps$ is decreased isothermally (thus decreasing
$\eps/T$).  This is continued until the entropy has increased by the
exact same amount $\Delta S$ lost in the first step.  Correspondingly
the system also withdraws an amount of heat $q_2=\Delta S/T_2$ from
this reservoir.  Finally the cold reservoir is removed and the
population inverted adiabatically once again to return to the upper
left corner.  Since the system withdraws heat from both reservoirs,
the conventionally defined Carnot efficiency $\eta=1-T_2/T_1$ is
larger than one, as claimed in the general discussion.

One point needs addressing since $T\propto\eps$ and population
inversion is achieved by reversing the sign of~$\eps$.  Clearly, this
avoids the aforementioned technical problem of adiabatically
connecting regions of opposite temperature through $1/T=0$ but it
means that the system passes through a state where $T=0$, in apparent
contravention of the third law of thermodynamics.  However, by keeping
$\eps/T$ finite, we are not putting all the spins in the same state,
which would be a violation of the third law.  Our thought experiment
violates no fundamental principle.  Nevertheless it is clear that any
attempt at a practical realization would be frustrated by the third
law, in the form which states that the entropy $S\to0$ as $T\to0$.
According to this, it is impossible to connect states with entropies
$S>0$ by an adiabatic process that passes through absolute zero.

The discrepancy arises, of course, because an idealized system of
non-interacting spins does not exist in reality.  Remnant interactions
provide the necessary means to enforce the third law in the real
world.\cite{Wal85} We can escape this by allowing a little
irreversibility to creep in, and ``jump around'' the problem at $T=0$.
This point has been discussed recently in \Refcite{Rom13}.  If
we do so, the efficiency will fall below that of the ideal, reversible
cycle.  Of course that will happen anyway in any practical
realization, but it seems to us there is no fundamental reason why the
efficiency of a practical engine could not also be bigger than one,
even if it falls short of the Carnot ideal.

\section{Discussion}
Let us summarize.  Our claim is that the Gibbs entropy fails to meet a
very basic expectation---the second law of thermodynamics.  The
Boltzmann entropy does meet this requirement.  Moreover, in the
thermodynamic limit the Boltzmann entropy also satisfies the
consistency criterion demanded by Dunkel and Hilbert.  Therefore we
see no reason why the Boltzmann entropy should be displaced from its
position as the lynch-pin connecting statistical mechanics with
thermodynamics.

Proponents of the Gibbs entropy may claim that we have taken it out of
context, and that it should only be discussed in relation to isolated
systems (\ie\ in the micro-canonical ensemble). This would seem to be a
singularly narrow definition of temperature since it means we are not
allowed to use it in the context of thermal equilibrium.\cite{Note6}
But, as we have seen for the three level spin system, even with such
an extreme position, the Gibbs entropy provides little in the way of
added value.

What are we to make of the fact that the Gibbs entropy satisfies
certain exact mathematical theorems, as adduced by Dunkel and Hilbert.
Of course we do not dispute these.  Rather, we say that for small
systems they are evidence of the well-known inequivalence of
ensembles, and the difficulty in finding a suitable entropy
definition.\cite{Nau05} For large systems, the theorems can be used to
prove that the Boltzmann entropy acquires certain desired features.
But if the exact theorems lead towards nonsensical conclusions, for
example that the temperature diverges exponentially with system size,
then what they are telling us is that the \emph{interpretation} is
wrong.  The original position is seen to become untenable.

With our viewpoint, negative temperatures are inevitable in systems
with bounded energy spectra.  The necessary extension of the formalism
of thermodynamics to treat this was described by Ramsey in
1956.\cite{Ram56} One must abandon the postulate that the entropy must
be an increasing function of energy.  But as so many times in the
history of science, an abandoned postulate opens the door to an
enriched formalism, capable of describing phenomena which were
not envisaged in the foundational era of the subject.

\section{Acknowledgments} 
We gratefully acknowledge feedback from Dick Bedeaux, Gavin Crooks,
Fabien Paillusson and Nicholas Tito.  We thank Roberto Piazza for
drawing our attention to the article by Berdichevsky,
\etal,~\cite{DH14} and Jose Vilar for sharing a relevant
preprint.\cite{VR14} We thank the authors of \Refcite{DH14}
for useful correspondence.

\section{Appendices}
The derivations here use rather more advanced mathematics than we have
employed in the main text, it seems appropriate to confine them to
Appendices.  We do not claim much originality and indeed the material
can be found in many textbooks.\cite{Sch52, Hil56}
\subsection{Thermodynamic consistency of $S_B$}\label{sec:appsb}
We prove that in the thermodynamic limit $\SB$ satisfies the
consistency criterion specified by Dunkel and Hilbert in
\Refcite{DH14}.

We first prove that in the \emph{canonical} ensemble\cite{Hil56}
\begin{equation}
-\fix{T}{\frac{\partial F}{\partial \Amu}}=
-\Bigl\langle\frac{\partial H}{\partial\Amu}\Bigr\rangle_T\,.
\label{eq:c1}
\end{equation}
In this, $\Amu$ is some parameter in the Hamiltonian $H$, for example
the position of a wall.  The left hand side (LHS) is the generalized
\emph{thermodynamic} force, corresponding to this parameter.  The
right hand side (RHS) is the canonical ensemble average of the
generalized \emph{mechanical} force, corresponding to the same
parameter.  These forces are said to be conjugate to $\Amu$.  For
example, in standard thermodynamics $(p,V)$ form a well-known
conjugate pair, and in our spin system $(-N_1,\eps)$ form another
conjugate pair.\cite{NoteA2} \Eqref{eq:c1} is well known and forms the
basis of a wide variety of Monte-Carlo free energy sampling
methods.

The ensemble average in \Eqref{eq:c1} is given by
\begin{equation}
\langle\cdots\rangle_T=
\frac{\Tr[(\cdots)\,e^{-\beta H}]}{\Tr[e^{-\beta H}]}\,,
\end{equation}
where ``$\Tr$'' can be read as ``sum over states'' and $\beta=1/\kB T$.
Provided there is a bounded energy spectrum, the sums remain
well defined for $\beta<0$.

The proof of \Eqref{eq:c1} is quite easy and starts from the expression
which defines the free energy,
\begin{equation}
e^{-\beta F}=\Tr[e^{-\beta H}]\,.\label{eq:bf}
\end{equation}
Differentiating both sides with respect to $\Amu$ gives
\begin{equation}
-\beta\,\fix{T}{\frac{\partial F}{\partial\Amu}}
e^{-\beta F}=-\beta\,\Tr\Bigl[\frac{\partial H}{\partial \Amu}
e^{-\beta H}\Bigr]\,.\label{eq:bdf}
\end{equation}
\Eqref{eq:c1} is obtained by dividing \Eqref{eq:bdf} by \Eqref{eq:bf},
and canceling $\beta$.

In order to recover the consistency criterion of Dunkel and Hilbert,
what we need to do is transfer these results to the
\emph{micro-canonical} ensemble.  To do this we first define the
micro-canonical ensemble average
\begin{equation}
\langle\cdots\rangle_E=\frac{\Tr[(\cdots)\,\delta(E-H)]}{\Tr[\delta(E-H)]}\,,
\end{equation}
and the micro-canonical Boltzmann entropy \via\
\begin{equation}
\Tr[\delta(E-H)]=e^{\SB/\kB}\,.
\end{equation}
In these the Dirac $\delta$-function selects only those states with
energy $E$.

With these in hand we can write 
\begin{equation}
\Tr[(\cdots)\,e^{-\beta H}]={\textstyle\int_0^\infty\!dE}\,
e^{-\beta E}\Tr[(\cdots)\,\delta(E-H)]
\end{equation}
and therefore
\begin{equation}
e^{-\beta F}\langle\cdots\rangle_T={\textstyle\int_0^\infty\!dE}\,
e^{-\beta E+\SB/\kB}\langle\cdots\rangle_E\,.\label{eq:dots}
\end{equation}
A special case obtains when $(\cdots)=1$, namely
\begin{equation}
e^{-\beta F}={\textstyle\int_0^\infty\!dE}\,
e^{-\beta E+\SB/\kB}\,.\label{eq:fb}
\end{equation}
\Eqsref{eq:dots} and~\eqref{eq:fb} have the structure of \emph{Laplace
  transforms}, with $(E, \beta)$ being the transform variables.  They
are valid irrespective of system size.\cite{NoteA2.5} As the system
size increases though, the integrand in both cases becomes dominated
by the peak in the exponential.  Therefore we can evaluate the
integrals by the saddle-point method.\cite{Sch52} Taking \Eqref{eq:fb}
first, one has
\begin{equation}
  -\beta F=-\beta E+\frac{\SB}{\kB}\quad\text{where}\quad
  -\beta+\frac{\partial}{\partial E}\Bigl(\frac{\SB}{\kB}\Bigr)=0\,.
  \label{eq:steep}
\end{equation}
The second expression is the condition that the integrand is a
maximum, as a function of $E$.  We now recall that $\beta=1/\kB T$ and
so the above can be rearranged to
\begin{equation}
  F=E-T\SB\quad\text{where}\quad 
  \frac{1}{T}=\frac{\partial\SB}{\partial E}\,.\label{eq:aa}
\end{equation}
Thus the saddle-point method has converted the Laplace transform into
a \emph{Legendre transform}.  \Eqref{eq:aa} looks very familiar but we
should emphasize that $T$ is the temperature in the canonical
ensemble.  So we have also proved that, in the thermodynamic limit,
\begin{equation}
  T=\TB\,.\label{eq:ttb}
\end{equation}
This can be viewed as an expression of the zeroth law of
thermodynamics.

Turning now to \Eqref{eq:dots} and proceeding in the same way, the
thermodynamic functions cancel, leaving
\begin{equation}
  \langle\dots\rangle_T=\langle\dots\rangle_E\,.\label{eq:ee}
\end{equation}
This confirms that ensemble averages are equivalent in the
thermodynamic limit.

The next step in the proof consists in differentiating the first of
\Eqsref{eq:aa} with respect to $\Amu$, paying careful attention to the
dependent variables.  One finds
\begin{equation}
  \fix{T}{\frac{\partial F}{\partial \Amu}}=\Bigl(1-T\,
  \frac{\partial\SB}{\partial E}\Bigr)\frac{\partial E}{\partial \Amu}
  -T\,\fix{E}{\frac{\partial\SB}{\partial \Amu}}\,.
\end{equation}
The first term on the RHS vanishes, by virtue of the second of
\Eqsref{eq:aa}.  Therefore
\begin{equation}
  \frac{\partial F}{\partial \Amu}\Big|_T=
  -T\,\frac{\partial\SB}{\partial \Amu}\Big|_E\,.
\end{equation}
Recalling \Eqsref{eq:c1}, \eqref{eq:ttb} and \eqref{eq:ee}, we
have now proved
\begin{equation}
  \TB\,\fix{E}{\frac{\partial\SB}{\partial \Amu}}=-\Bigl\langle\frac{\partial
    H}{\partial \Amu}\Bigr\rangle_E\,.\label{eq:c2}
\end{equation}
This is exactly Dunkel and Hilbert's consistency criterion
(\cf\ Eq.~(7) in \Refcite{DH14}).  It holds in the
thermodynamic limit and is our desired result.\cite{NoteA3}

\subsection{Some properties of $\TG$}\label{sec:apptg}
We wish to know the behavior of $\TG$ in the limit of large system
size.  Starting from the definitions, $\kB \TG=
(\partial\ln\Omega/\partial E)^{-1}$ and $\partial\Omega/\partial E =
\omega$, we find $\kB \TG=\Omega/\omega$, as in
\Refcite{DH14}. From this we obtain
\begin{equation}
\TG=\kB^{-1}{\textstyle\int_0^E\!dE'}\, e^{[\SB(E')-\SB(E)]/\kB}\,.
\label{eq:a2a}
\end{equation}
This compact expression essentially contains everything we need to
know about the behavior of $\TG$.  The argument of the exponential is
proportional to the system size (\ie\ extensive).  Therefore, for
large systems, the value of $\TG$ is determined by the location of the
maximum.  

There are two cases.  If $\SB$ is monotonically increasing
up to $E$, the integral is determined by the behavior at the upper
limit of integration.  Setting $\Delta E=E-E'$ we first have
\begin{equation}
\SB(E')-\SB(E)=-{\Delta E}/{\TB}+O(\Delta E^2)
\end{equation}
since $1/\TB=\partial\SB/\partial E$.  Changing integration variable
to $\Delta E$ we then have 
\begin{equation}
\TG=\kB^{-1}{\textstyle\int_0^E\!d(\Delta E)}\,e^{-\Delta E/\kB\TB}\,.
\end{equation}
The upper integration limit can be replaced by $\infty$ since the
correction becomes vanishingly small in the limit of a large system
size.  The integral can then be done,
\begin{equation}
\TG=\kB^{-1}{\textstyle\int_0^\infty\!d(\Delta E)}\, e^{-\Delta E/\kB\TB}=\TB\,.
\end{equation}
Therefore we conclude that as long as $\SB$ is monotonically
increasing, $\TG=\TB$ in the thermodynamic limit.

In the second case the maximum of the argument of the exponential in
Eq.~\eqref{eq:a2a} occurs at $E'=\Estar$, somewhere between the
integration limits (note that $\SB(E)$ is just an offset).
Approximating the integral by the maximum value of the integrand we
find
\begin{equation}
\TG=\kB^{-1}\,e^{[\SB(\Estar)-\SB(E)]/\kB}\,.
\label{eq:a2b}
\end{equation}
This demonstrates that to leading order $\TG$ grows exponentially with
system size.\cite{VR14} Again recalling that
$1/\TB=\partial\SB/\partial E$, we have finally
\begin{equation}
\TG=\kB^{-1}\exp(-{\textstyle\int_\Estar^E dE'}/\kB\TB)\,.
\label{eq:a2c}
\end{equation}
Hence to leading order $\TG$ acquires a complicated dependence on the
Boltzmann temperature between $\Estar$ and $E$.  Note
that $\TB<0$ in \Eqref{eq:a2c} so $\TG$ is still an increasing
function of $E$.


\begin{thebibliography}{99}
\bibitem{Note4.1} Excepting that a minor improvement is required
  in the Kelvin-Planck formulation of the second law.\cite{Zem57,
    Lan77}
\bibitem{Note4.2} Writing $dF=(N_1/N)\, d\Emax$ makes this precise:
  $-N_1/N$ is the analogue of $p$, and $\Emax=\eps N$ is the analogue
  of $V$.
\bibitem{NoteA2} In the spin system, the Hamiltonian
  $H=\eps\sum_{i=1}^M s_i$, where $s_i$ is 0 or 1 according to whether
  the $i$-th spin is in the ground state or excited state.
  \Eqref{eq:c1} is then satisfied almost trivially since
  $\langle\partial H/\partial\eps\rangle=\langle\sum_{i=1}^N
  s_i\rangle=N_1$.
\bibitem{NoteA2.5} This demonstrates the well known ensemble
  inequivalence for small systems since in that case $-\partial
  F/\partial T\ne\SB$.
\end{thebibliography}
\end{document}